\documentclass[twocolumn,showpacs,preprintnumbers,amsmath,amssymb]{revtex4}

\usepackage{graphicx}
\usepackage{dcolumn}
\usepackage{bm}

\newcommand\beq{\begin{eqnarray}}
\newcommand\eeq{\end{eqnarray}}

\def\kvec{\mbox{\boldmath $k$}_\perp}
\def\k3vec{\mbox{\boldmath $k$}}
\def\bvec{\mbox{\boldmath $b$}}
\def\lvec{\mbox{\boldmath $l$}_\perp}

\def\Delvec{\mbox{\boldmath $\Delta$}_\perp}
\def\0vec{\mbox{\boldmath $0$}_\perp}

\def\vep{\varepsilon}

\def\slash#1{\rlap/{#1}}

\def\vepslash{\slash{\mkern-1mu \varepsilon}}
\def\eslash{\slash{\mkern-1mu e}}

\def\qtilslash{\slash{\mkern-1mu \tilde{q}}}

\def\es{s}

\input{epsf.sty}
\input{epsfig.sty}

\begin{document}


\title{
Diffractive $\eta_{c}$ and $\eta_{b}$
productions 
by neutrinos
via 
neutral currents}

\author{A. Hayashigaki$^{(1)}$, K. Suzuki$^{(2)}$ and K. Tanaka$^{(3)}$}
\address{$^{(1)}$ Department of Physics, University of Tokyo, Tokyo 113-0033, 
Japan}\email{arata@nt.phys.s.u-tokyo.ac.jp, arata.hayashigaki@physik.uni-regensburg.de}%
\address{$^{(2)}$ Division of Liberal Arts, Numazu College of Technology, Shizuoka 410-8501, 
Japan}\email{ksuzuki@la.numazu-ct.ac.jp}%
\address{$^{(3)}$ Department of Physics, Juntendo University, Inba-gun, 
Chiba 270-1695, Japan}\email{tanakak@sakura.juntendo.ac.jp}%

\date{\today}

\begin{abstract}
We report a first theoretical study for neutrino-induced 
diffractive productions of heavy pseudoscalar mesons, 
$\eta_c$ and $\eta_b$, off a nucleon. 
Based on factorization formalism for exclusive processes,
we evaluate the forward diffractive production cross section 
in perturbative QCD in terms of the light-cone $Q\bar{Q}$ wave functions (WFs) of 
$\eta_{c, b}$ mesons
and the gluon distribution of the nucleon.
The light-cone WFs of the $\eta_c$ ($\eta_{b}$) meson 
are constructed to satisfy the spin symmetry
relations with those of the $J/\psi$ ($\Upsilon$) meson.
The diffractive $\eta_c$ production 
is governed by the axial-vector coupling of the longitudinally polarized $Z$ boson
to $Q\bar{Q}$ pair, and 
the 
resulting 
$\eta_c$ production rate
is larger than the $J / \psi$ one
by one order of magnitude.
We also discuss the production of 
bottomonium $\eta_b$, which shows up for higher beam energy. 
\end{abstract}

\pacs{12.38.Bx, 12.39.St, 13.60.Le, 14.70.Hp}

\maketitle



Exclusive diffractive leptoproductions of neutral vector mesons 
provide
unique insight into an interplay between nonperturbative and perturbative 
effects in QCD \cite{Levin}.
The diffractive processes are mediated 
by the exchange of a Pomeron with the vacuum quantum numbers,
whose QCD description is directly related to the gluon distributions inside the 
nucleons for small Bjorken-$x$ \cite{Ry,BFGMS,FKS,RRML,LMRT}.
The processes also allow us 
to probe the light-cone wave functions (WFs)
of the vector mesons.
Relating to the latter point, however, the applicability is 
apparently limited to probing the neutral vector mesons.

\begin{figure}[h]
\psfig{file=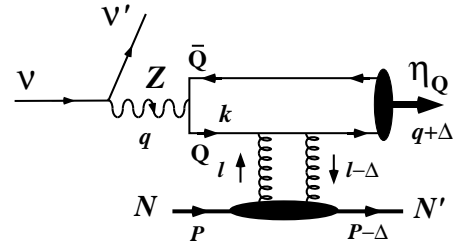,width=6cm}
\caption{
A typical diagram for the exclusive diffractive $\eta_Q$ ($Q=c,b$) productions 
induced by neutrino ($\nu$) through the $Z$ boson exchange. 
There are other diagrams by interchanging the vertices
on the heavy-quark lines.}
\label{fig:etac_prod2}
\end{figure}

In this Letter, we propose the exclusive diffractive productions of mesons 
in terms of the neutrino beam. 
The weak currents allow us to observe both neutral ($M^{0}$)
and charged mesons ($M^{\pm}$) as 
$\nu_\mu + N \to \nu_\mu / \mu^\mp  + N + M^0 / M^\pm$ 
by $Z/W$ boson exchange \cite{KM}, and these mesons
can be not only vector but 
also other types of mesons including pseudoscalar mesons.
Thus, such processes may reveal structure of various kinds of mesons, 
the coupling of the QCD Pomeron to quark-antiquark pair with various spin-flavor
quantum numbers, and 
information on the CKM matrix elements.  
There 
already 
exist some experimental data for 
$\pi$, $\rho$, $D_s^{\pm}$, $D_s^{*}$ \cite{E632,E815}, 
$D_s^{*+}$ \cite{CHORUS}, and $J / \psi$ production \cite{E815}, 
but there are only a few theoretical calculations, {\it e.g.},
for the $J / \psi$ production in a vector meson dominance model \cite{BKP}
and for $D_s^-$ production 
with the generalized parton density \cite{LS}. 
Our work gives a first 
QCD calculation for diffractive meson productions via the weak neutral current.
Here, 
our interest will be directed to 
productions of heavy pseudoscalar mesons, especially 
$\eta_c$ and $\eta_b$ as in Fig.~\ref{fig:etac_prod2}.
So far $\eta_c$ 
has been observed via the decays of $J/\psi$ or $B$ mesons
produced by $p\bar{p}$ and $e^+e^-$ reactions, while 
$\eta_b$ has not been observed.
The diffractive productions via the weak neutral current
will give a direct access to $\eta_{c}$ as well as a new 
experimental method to identify 
$\eta_b$ by {\it e.g.} measuring the two photon decay.

We treat the $\eta_c$ and $\eta_b$ productions by 
generalizing the approach in the leading logarithmic order of perturbative QCD, 
which has been developed 
successfully for the vector meson
electroproductions \cite{Ry,BFGMS,FKS,RRML,LMRT}.  
We consider the near-forward diffractive productions 
$Z^* (q)+ N (P) \to \eta_Q (q+\Delta) + N' (P-\Delta)$,
where $Q=c, b$, and $q, q+\Delta, P$ and $P-\Delta$ denote the momenta
of the virtual $Z$ boson, $\eta_{Q}$ meson, initial and final nucleons, respectively.
The total center-of-mass energy 
$W= \sqrt{(P + q)^2}$ is much larger than 
any other mass scales involved, i.e.,
$W^2 \gg K^{2}$ with $K^{2} = {\cal Q}^{2}, -t, m_{Q}^{2}, \Lambda_{\rm QCD}^2, \ldots$, where 
${\cal Q}^2 =-q^2$, $t= \Delta^2$, and 
$m_{Q}$ is the heavy-quark mass.
We also suppose $- t \ll m_{Q}^2$ and $\Lambda_{\rm QCD}^2\ll m_{Q}^2$.
%
%
The heavy quark mass $m_Q$ ensures that perturbative QCD can be
applied even for ${\cal Q}^2 =0$ to calculate the creation of the $Q\bar{Q}$ pair as well as
its time development before the nonperturbative effects to form a quarkonium state $\eta_Q$ set in. 
The crucial point is 
that at high $W$ the scattering of the $Q\bar{Q}$ pair on the nucleon
occurs over a much shorter timescale than the $Z^{*} \rightarrow Q\bar{Q}$
fluctuation or the 
$Q\bar{Q} \rightarrow \eta_Q$
formation times (see Fig.~\ref{fig:etac_prod2}).
As a result, the production amplitudes obey factorization 
in terms of the $Z$ and $\eta_{Q}$ light-cone WFs.
The $Q \bar Q$-$N$ elastic scattering amplitude, 
sandwiched between the  $Z$ and $\eta_{Q}$ WFs,
is further factorized into the $Q \bar Q$-gluon hard scattering amplitude 
and the nucleon matrix element corresponding to the (unintegrated) gluon density distribution.
This latter step in the factorization to get the gluon distribution 
can be carried out in the same manner as in the previous works \cite{Ry,BFGMS,FKS,RRML,LMRT}.
The participation of 
the ``new players'' $Z$ and $\eta_Q$ in the former step 
requires an extension
of the previous works
by introducing the corresponding light-cone WFs.

%

First of all,
we discuss the extension due to the participation of the $Z$ boson.
The $ZQ\bar{Q}$ weak vertex of Fig.~\ref{fig:etac_prod2}
is given by 
$(g_W/2\cos \theta_W)\gamma_\mu (c_V-c_A\gamma_5)$,
where $(g_{W}/2 \cos \theta_W)^2 = \sqrt{2}G_F M_Z^2$ with $G_F$ the Fermi
constant and $M_Z$ the $Z$ mass. 
$c_V=1/2 - (4/3) \sin^2 \theta_W, c_A = 1/2$ for the $c$-quark and similarly 
for the $b$-quark.
As usual,
we introduce the two light-like vectors 
$q'$ and $p'$ by the relations
$q=q'-({\cal Q}^2/\es)p'$, $P=p' + (M_{N}^{2}/\es)q'$,
$q^{\prime 2} =p^{\prime 2} =0$, $\es=2 q'\cdot p'$
with $M_N$ the nucleon mass and $\es \cong  W^{2}+{\cal Q}^{2}-M_{N}^{2}$,
and the Sudakov decomposition of all momenta,
{\it e.g.}, $k=\alpha q' + \beta p' + k_{\perp}$.
\begin{figure}[h]
\psfig{file=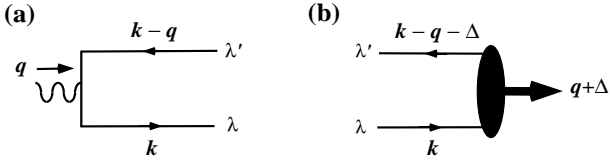,width=9cm}
\caption{Diagrams
representing the light-cone WFs: (a) for the virtual $Z$ boson; (b)
for the $\eta_{Q}$ meson.}
\label{fig:zwf}
\end{figure}
We also introduce 
the polarization vectors $\vep^{(\xi)}$ ($\xi = 0, \pm 1$) of the virtual $Z$ boson to satisfy
$\sum_{\xi}(-1)^{\xi + 1}{\varepsilon_{\mu}^{(\xi)}}^{*}\varepsilon_{\nu}^{(\xi)}
= - g_{\mu \nu} + q_{\mu}q_{\nu}/M_{Z}^{2}$,
which is the numerator of the propagator 
for the massive 
vector boson. Because the 
$q_{\mu}q_{\nu}/M_{Z}^{2}$ term vanishes when contracted with the neutral current,
we conveniently choose as
\begin{eqnarray} 
\vep^{(0)} = \frac{1}{{\cal Q}}q'+\frac{{\cal Q}}{\es}p', \;\;\;\;
\vep^{(\pm 1)} = \vep^{(\pm 1)}_{\perp} = \frac{1}{\sqrt{2}}(0,1,\pm i,0) ,
\label{eqn:2-6}
\end{eqnarray}
for the longitudinal ($\xi =0$) and transverse ($\xi = \pm 1$) polarizations, respectively.
Eq.~(\ref{eqn:2-6}) coincides with the polarization vectors
for the virtual photon in Refs. \cite{Ry,BFGMS,FKS,RRML}.
For the $J/\psi$ electroproductions,
the wavy line of Figs.~\ref{fig:etac_prod2} and \ref{fig:zwf}(a) 
denotes a virtual photon.
The corresponding photon 
light-cone WFs 
are derived as \cite{BFGMS,FKS,GQ}
\begin{eqnarray}
\Psi^{\gamma (\xi)}_{\lambda \lambda'} (\alpha,\kvec) = - e_{Q}
\frac{\sqrt{\alpha(1-\alpha)} \bar{u}_{\lambda}(k) \vepslash^{(\xi)}
v_{\lambda'}(q-k)}
{\alpha (1-\alpha) {\cal Q}^2 + \kvec^2 +m_Q^2},
\label{eqn:phwf}
\end{eqnarray}
with eq.~(\ref{eqn:2-6}) for $\xi=0, \pm 1$.
Here $e_{Q}$ is the electric charge of the quark, and 
$u_{\lambda}(k)$ ($v_{\lambda'}(q-k)$) denote the on-shell spinor for the (anti)quark
with helicity $\lambda$ ($\lambda'$).
These WFs describe the probability amplitudes to find the photon
in a state with the $Q\bar{Q}$ pair, where the quark carries the longitudinal momentum fraction 
$\alpha$ and the transverse momentum $\kvec$ as $k = \alpha q'+\beta p'+k_\perp$,
$k_\perp^2 = - \kvec^2$.
Obviously, 
the light-cone WFs
for the virtual $Z$ boson can be obtained from eq.~(\ref{eqn:phwf}) 
by the replacement $e_{Q} \gamma_{\mu} \rightarrow 
(g_W/2\cos \theta_W)\gamma_\mu (c_V-c_A\gamma_5)$ as 
\begin{eqnarray}
\lefteqn{\Psi^{Z(\xi)}_{\lambda \lambda'} (\alpha,\kvec) = - \frac{g_W}{2 \cos \theta_W}}
\nonumber\\
&&\times
\frac{\sqrt{\alpha(1-\alpha)} \bar{u}_{\lambda}(k) \vepslash^{(\xi)}
\left(c_{V} - c_{A}\gamma_{5}\right)v_{\lambda'}(q-k)}
{\alpha (1-\alpha) {\cal Q}^2 + \kvec^2 +m_Q^2}.
\label{eqn:2-12}
\end{eqnarray}

Next 
we proceed to the light-cone WFs for the $\eta_{Q}$ meson,
corresponding to Fig.~\ref{fig:zwf}(b).
Again, it is convenient to exploit the correspondence with
the $J/\psi$ electroproductions.
The light-cone WFs for the heavy vector
mesons $V=J/\psi, \Upsilon$ have been discussed in many works,
but are still controversial 
in the treatment of subleading
effects like the Fermi motion corrections \cite{FKS,RRML,hood}, corrections to ensure
the pure $S$-wave $Q\bar{Q}$ state \cite{ivanov}, corrections via the Melosh rotation \cite{HIKT}, {\it etc}.
Here we employ the light-cone WFs for the vector meson 
given by (see Fig.~\ref{fig:zwf}(b))
\begin{eqnarray}
{\Psi^{V(\varpi)}_{\lambda \lambda'}}^{*}\!\!
(\alpha,\mbox{\boldmath $k$}_\perp)
\!=\! \frac{\bar{v}_{\lambda'}(\tilde{q}-k)}{\sqrt{1-\alpha}}\gamma^{\mu}
{e_{\mu}^{(\varpi)}}^{*} {\cal R}
\frac{u_{\lambda}(k)}{\sqrt{\alpha}} 
\frac{\phi^*(\alpha,\kvec)}{M_{V}} , 
\label{eqn:V}
\end{eqnarray}
where $\tilde{q} = q+ \Delta$, 
$M_{V}$, and $e_{\mu}^{(\varpi)}$ ($\varpi = 0, \pm 1$) are momentum, mass, 
and the polarization vector of the vector meson
with $\tilde{q}^2 = M_{V}^{2}$, $e^{(\varpi)}\cdot \tilde{q} = 0$, and 
$e^{(\varpi) *}\cdot e^{(\varpi')}= - \delta_{\varpi \varpi'}$.
${\cal R} \equiv (1+\qtilslash/M_{V})/2$ denotes the projection operator,
${\cal R}^2 = {\cal R}$,   
to ensure the $S$-wave $Q\bar{Q}$ state in the heavy-quark limit \cite{Ry,hood,BJ}.
(This projection operator coincides with that discussed in Ref.~\cite{ivanov}
up to the binding-energy effects of the quarkonia.)
Note that eq.~(\ref{eqn:V}) reduces to the vector-meson WFs of Ref.~\cite{BFGMS}
by the replacement ${\cal R} \rightarrow 1$.
The physical interpretation of eq.~(\ref{eqn:V}) is similar to the ``perturbative'' WFs
(\ref{eqn:phwf}) and
(\ref{eqn:2-12}), but the scalar function $\phi(\alpha,\kvec)$ contains 
nonperturbative dynamics between $Q$ and $\bar{Q}$. 
Now, the light-cone
WFs for the $\eta_{Q}$ meson 
can be derived from eq.~(\ref{eqn:V}) utilizing spin symmetry,
which is exact in the heavy-quark limit. 
This symmetry relates the $S$-wave states, $\eta_{Q}$ and
the three spin states of the vector meson \cite{Geo}.
Namely, $M_{\eta_Q} = M_{V}$, and 
the pseudoscalar state is related to the vector state with
longitudinal polarization as
$|\eta_{Q} \rangle = 2 \hat{S}_{Q}^{3} |V(\varpi=0)\rangle$,
where 
$\hat{S}_{Q}^{3}$ is the third component of
the hermitean spin operator $\hat{S}_{Q}^{i}$ 
which acts on the spin of the heavy quark $Q$ but does not act on $\bar{Q}$.
For the WFs we get (see eq.~(\ref{eqn:V}))
\begin{eqnarray}
\Psi^{\eta_{Q} *}_{\lambda \lambda'}
(\alpha,\mbox{\boldmath $k$}_\perp)
&&\!\!\!\!\!\!\!
=\! \frac{\bar{v}_{\lambda'}(\tilde{q}-k)}{\sqrt{1-\alpha}}\gamma^{\mu}
{e_{\mu}^{(0)}}^{*} {\cal R} \left( 2 S^{3} \right)
\frac{u_{\lambda}(k)}{\sqrt{\alpha}} 
\frac{\phi^*(\alpha,\kvec)}{M_{V}} \nonumber\\
=-&&\!\!\!\!\! \frac{\bar{v}_{\lambda'}(\tilde{q} -k)}{\sqrt{1-\alpha}}\gamma_5
{\cal R}
\frac{u_{\lambda}(k)}{\sqrt{\alpha}} \ 
\frac{\phi^*(\alpha,\kvec)} {M_{\eta_Q}} .
\label{eqn:3-1}
\end{eqnarray}
Here $S^{3}$ is a matrix repesentation of $\hat{S}_{Q}^{3}$
as $S^{3} = \gamma_{5} \qtilslash \eslash^{(0)}/(2M_{V})$,
which is related to a spin matrix $\sigma^{12}/2=\gamma_{5}\gamma^{0}\gamma^{3}/2$
in the meson rest frame
by a Lorentz boost in the third direction \cite{Geo}.
Thus the $\eta_{Q}$ is described by the same 
nonperturbative WF 
$\phi(\alpha,\kvec)$ as the vector meson.
Note that, due to the presence of ${\cal R}$,
the ``$Q\bar{Q}\eta_{Q}$ vertex'' involves pseudovector  
as well as pseudoscalar coupling.

Combining our $Z$ and $\eta_{Q}$ WFs 
with the $Q\bar{Q}$-$N$ elastic amplitude which was obtained in Ref.~\cite{BFGMS},
we get the total amplitude for the polarization $\varepsilon^{(\xi)}$ of the 
$Z$ boson as
\begin{eqnarray}
\mbox{Im}{\cal M}^{(\xi)}
&=& \frac{\sqrt{2}\pi s}{2\sqrt{N_c}}\ \sum_{\lambda'\lambda}
\int \frac{d\alpha d^2 k_{\perp}}{16\pi^3}\int d^{2}l_{\perp}\ \frac{\alpha_s(\lvec^2)}
{\lvec^4}
\nonumber\\
\times&&\!\!\!\!\!\!\!\!\!\!
f (x,\lvec^2)
\left[\,2\Psi^{Z(\xi)}_{\lambda \lambda'}(\alpha,\kvec)-\Psi^{Z(\xi)}_{\lambda \lambda'}(\alpha,\kvec+\lvec)
\right. \nonumber\\
-&&\!\!\!\!\!\!\!\!\!\!
\left. \Psi^{Z(\xi)}_{\lambda \lambda'}(\alpha,\kvec-\lvec)\,\right]\
{\Psi^{\eta_Q}_{\lambda \lambda'}}^*(\alpha,\kvec) ,
\label{eqn:4-4}
\end{eqnarray}
%
up to the terms suppressed for high $W$ and $\Delvec^2 \cong -t \rightarrow 0$.
Here $\alpha_{s}(\lvec^{2})$ denotes the running coupling constant
for $N_{c}$ colors.
As usual, 
we have explicitly dealt with the imaginary part of the production amplitude, because 
the small real part can be reconstructed perturbatively 
(see eq.~(\ref{eqn:5-1}) below).
$f(x,\lvec^2)$ denotes the unintegrated gluon density,
and $x= ({\cal Q}^2+M_{\eta_Q}^2)/\es$ and $\lvec$ are the longitudinal momentum fraction
and the transverse momentum, respectively, which are
carried by a gluon inside the nucleon \cite{BFGMS,RRML}.
We substitute eqs.~(\ref{eqn:2-12}) and (\ref{eqn:3-1})
into eq.~(\ref{eqn:4-4}),
and go over to the ``$\bvec$-space''
conjugate to the $\kvec$-space via the 
Fourier transformation
$\phi(\alpha, \kvec) = 4\pi \int d^2 b e^{-i \kvec \cdot \bvec} \phi(\alpha, \bvec)$; 
$\bvec$ denotes the transverse separation between $Q$ and $\bar{Q}$. 
Then, it is straightforward to see $\mbox{Im}{\cal M}^{(\pm 1)} = 0$
which reflects helicity conservation 
in the high energy limit.
$\mbox{Im}{\cal M}^{(0)}$ can be calculated 
in parallel with previous works for the vector meson
production \cite{RRML,FKS,SHIAH}.
In the integrand, there appear the terms $e^{i\lvec\cdot \bvec}$ and $e^{-i\lvec\cdot \bvec}$.  
The leading $\ln ({\cal Q}^2/\Lambda_{\rm QCD}^{2})$ contribution comes from
the region $\Lambda_{\rm QCD}^{2} \ll \lvec^2 \ll {\cal Q}^2_{\rm eff}(\equiv [{\cal Q}^2+M_{\eta_Q}^2]/4)$
of the $\lvec$-integral \cite{BFGMS,RRML}, 
where $\lvec \cdot \bvec \ll 1$ is satisfied because $|\bvec|\sim 1 / m_Q$.
We retain only the leading nonzero term in 
the power series in 
$\lvec\cdot \bvec$,
which corresponds to the ``color-dipole picture'' \cite{FKS,SHIAH}.
Then, using 
$\int^{{\cal Q}^{2}_{\rm eff}}  dl_{\perp}^{2}  f(x, \lvec^{2})/\lvec^{2} 
= xG(x, {\cal Q}_{\rm eff}^{2})$ 
with $G(x, {\cal Q}_{\rm eff}^{2})$ the conventional gluon distribution,
we get ($b\equiv |\bvec|$) \cite{SHIAH}
\begin{eqnarray}
i{\cal M}^{(0)}
&&
\!\!\!\!\!\!\!
=\frac{-\sqrt{2}\pi^2 W^2}{\sqrt{N_c}}\frac{g_W m_Q c_A}{M_{\eta_Q}{\cal Q}\cos \theta_W}
\alpha_s({\cal Q}_{\rm eff}^2)\!
\left[1+i\frac{\pi}{2}\frac{\partial}{\partial 
\mbox{ln}x}\right]
\nonumber\\
\times
x&&
\!\!\!\!\!\!\!
G(x,{\cal Q}_{\rm eff}^2)
\!\!
\int_0^1 
\!\!\! 
\frac{d\alpha{\cal Q}_{m}}{\alpha(1-\alpha)}
\int_0^\infty  
\!\!\!\!\! 
dbb^2\phi^*(\alpha,b)
K_1\left(b{\cal Q}_{m}\right) ,
\label{eqn:5-1}
\end{eqnarray}
where
${\cal Q}_{m}=[\alpha(1-\alpha){\cal Q}^2+m_Q^2 ]^{1/2}$, 
$K_{1}$ is a modified Bessel function, and
we have included the real part of the amplitude 
as a perturbation \cite{BFGMS,FKS,RRML}.
As expected, the result (\ref{eqn:5-1}) is proportional to $c_A$, 
so that the $\eta_{Q}$ meson is generated by the axial-vector part of the 
weak current.

For comparison, we also calculate 
the diffractive 
vector meson production via the weak neutral current, by the replacement 
$\Psi^{\eta_{Q}}_{\lambda \lambda'} \rightarrow  \Psi^{V(\varpi)}_{\lambda \lambda'}$
in eq.~(\ref{eqn:4-4}). Substituting eq.~(\ref{eqn:V}), 
we find that $\xi=\varpi$, i.e., ${\cal M}^{(0)}$
and ${\cal M}^{(\pm 1)}$ give the production of the longitudinally and transversely
polarized vector mesons, respectively,
and that all these amplitudes
are proportional to the vector coupling $c_{V}$.
(When we replace the factor $g_{W} c_{V}/(2 \cos \theta_W )$ by $e_{c}=2 e /3$
in these results,
we obtain the amplitudes ${\cal M}^{(\xi)}$ identical to those for the diffractive 
electroproduction of $J / \psi$ \cite{FKS,RRML}, up to the corrections due to
${\cal R}$ of eq.~(\ref{eqn:V}).)

Combining eq.~(\ref{eqn:5-1}) with the $Z$ boson propagator and the weak neutral current by a neutrino, 
the forward differential cross section for the $\eta_{Q}$ production 
is given by
\begin{eqnarray}
\left.\frac{d^3\sigma(\nu N \rightarrow \nu' N' \eta_Q )}{d\es d{\cal Q}^2 dt}
\right|_{t=0}
=&&
\!\!\!\!
\frac{1}{4(8\pi)^3E_\nu^2 M_N^{2} \es}
\frac{g_W^2}{\cos \theta_W^2}
\nonumber\\
\times\mbox{\hspace{0.8cm}}&&
\!\!\!\!\!\!\!\!\!\!\!\!\!\!\!\!\!\!\!\!
\frac{{\cal Q}^2}{\left({\cal Q}^2+M_Z^2\right)^2}
\frac{\epsilon}{1-\epsilon}
\left|{\cal M}^{(0)}\right|^2,
\label{eqn:5-2}
\end{eqnarray}
where $E_{\nu}$ is the neutrino beam energy in the lab system, and 
$\epsilon = [4(1-y)-Q^2/E_\nu^2]/[2\{1+(1-y)^2\}+Q^2/E_\nu^2]$ 
with $y=\es /(2M_{N}E_\nu )$
is the polarization parameter of the virtual $Z$ boson \cite{KM,BKP}.
In order to evaluate 
the corresponding elastic $\eta_{Q}$ production rate,
we assume the $t$-dependence as 
$d^3\sigma/d\es d{\cal Q}^{2} dt =d^{3}\sigma/d\es d{\cal Q}^{2} dt|_{t=0}\, \exp (B_{\eta_{Q}} t)$ 
with a constant diffractive slope, as in the case of the vector meson production.
Integrating $d^2\sigma/d\es d{\cal Q}^{2}= (1/B_{\eta_{Q}})d^{3}\sigma/d\es d{\cal Q}^{2} dt|_{t=0}$ over 
${\cal Q}^{2}$ and $\es$,
we get the elastic production rate
$\sigma(\nu N \rightarrow \nu' N' \eta_Q )$
as a function of $E_\nu$.

For numerical computation of $\sigma( \nu N \rightarrow \nu' N' \eta_Q )$,
we need explicit form of 
$\phi(\alpha,\bvec)$ of
eq.~(\ref{eqn:5-1}).
From eqs.~(\ref{eqn:V}) and (\ref{eqn:3-1}),
we can use the corresponding nonperturbative part of the vector-meson WF
which was constructed in the previous works based on, {\it e.g.}, 
non-relativistic potential model for heavy quarkonia \cite{FKS,SHIAH}.
In this work, we adopt the WF from the Cornell potential model 
with the corresponding quark masses $m_c=1.5$ GeV, $m_b=4.9$ GeV \cite{EQ}.
A procedure to construct 
$\phi(\alpha,\kvec)$ with the light-cone variables
from the usual Schr\"{o}dinger WF $\phi_{NR} (\k3vec)$ 
has been given in Refs.~\cite{FKS,SHIAH}. 
Also, we use the empirical values for the masses $M_{\eta_c}=2.98$ GeV, 
$M_N=0.94$ GeV and $M_Z=91.2$ GeV.
For $\eta_{b}$, we use $M_{\eta_b}=9.45$ GeV estimated in Ref.~\cite{HK}. 
Because the slope $B_{\eta_{Q}}$ introduced above is unknown, 
we assume that $B_{\eta_{Q}}$ has the same value
as that for the corresponding vector meson:
$B_{\eta_{c}} =4.5$ GeV$^{-2}$ from the 
experimental value for $J / \psi$ \cite{RRML} and 
$B_{\eta_b}=3.9$ GeV$^{-2}$ following the value for $\Upsilon$ in Ref.~\cite{FMS}.
For the gluon distribution 
$G(x,{\cal Q}_{\rm eff}^2)$ 
of eq.~(\ref{eqn:5-1}),
we employ 
GRV95 NLO parameterization \cite{GRV95}.
\begin{figure}[htb]
\psfig{file=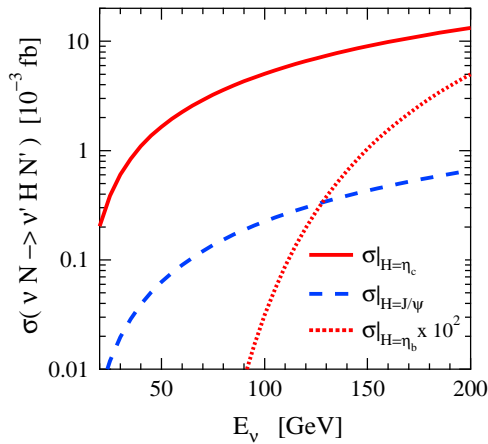,width=7cm}
\caption{The elastic production rates as functions of $E_\nu$.
Solid, dashed and dotted curves are for $\eta_c$, $J /\psi$ and $\eta_{b}$,
respectively. 
Note that the dotted curve shows the rate multiplied by $10^2$.}
\label{fig:tot_etac}
\end{figure}

We show the elastic $\eta_{c}$ production rate,
$\sigma( \nu N \rightarrow \nu' N' \eta_c )$,
by the solid curve in Fig.~\ref{fig:tot_etac}.
The result monotonically increases as a function of the beam energy $E_\nu$.
Such behavior is similar with that observed in the $\pi$-production data 
by the neutral \cite{KM} and charged currents \cite{E632}.
For comparison, we show the  
elastic $J/\psi$ production rate $\sigma( \nu N \rightarrow \nu' N' J/\psi )$
by the dashed curve
(see the discussion below eq.~(\ref{eqn:5-1})) \cite{foot}.
The rate for $\eta_c$ production 
is much larger than that for $J/\psi$
by a factor $\sim 20$.
This is mainly due to the relevant weak couplings,
$c_{A}$ for $\eta_{c}$ and $c_{V}$ for $J/\psi$, 
as $(c_{A}/c_{V})^{2} \cong 7$.
Another important effect comes from 
the different behavior of the $Z$ light-cone WFs (\ref{eqn:2-12}) 
between the axial-vector and vector 
channels:
It is straigtforward to see that the $\eta_{c}$ diffractive amplitude (\ref{eqn:5-1})
as well as the corresponding amplitudes for the longitudinally and
transversely polarized $J/\psi$ 
decreases rather rapidly for high ${\cal Q}^{2}$ region,
${\cal Q}^{2} \gtrsim 4m_{c}^{2}$, so that the elastic production rates 
are in principle dominated by the low ${\cal Q}^{2}$ contribution
of the differential cross sections 
(similar behavior
is observed also for $J/\psi$ electroproductions \cite{FKS,RRML}).
For weak neutral current proceses, however, a typical factor ${\cal Q}^2 /({\cal Q}^2 +M_Z^2 )^2$
appears in the diffrential cross sections as in eq. (\ref{eqn:5-2}).
The low ${\cal Q}^{2}$ contribution of the 
differential cross section for $J/\psi$ is suppressed by this factor,
while that for $\eta_{c}$ avoids the suppression due to $\sim 1/{\cal Q}$ behavior of eq. (\ref{eqn:5-1}).
This particular behavior of the $\eta_{c}$ diffractive amplitude comes from the 
spinor matrix element of the $Z$ light-cone WFs (\ref{eqn:2-12}) as
$\bar{u}_{\lambda}(k) \vepslash^{(0)}\gamma_{5} v_{\lambda'}(q-k) \sim 1/{\cal Q}$ for ${\cal Q} \rightarrow 0$.
On the other hand, $\bar{u}_{\lambda}(k) \vepslash^{(0)}v_{\lambda'}(q-k) \sim {\cal Q}$
and $\bar{u}_{\lambda}(k) \vepslash^{(\pm1)}v_{\lambda'}(q-k) \sim {\rm const}$, for the vector
channels relevant for $J/\psi$.




In Fig.~\ref{fig:tot_etac}, we also show the $\eta_{b}$ production rate
$\sigma( \nu N \rightarrow \nu' N' \eta_b )$
by the dotted curve.
Although the rate for $\eta_{b}$ is generally much
smaller than that for $\eta_{c}$,
the former increases more rapidly than the latter
for increasing $E_\nu$.
Therefore, the $\eta_{b}$ production rate
could become comparable with the $\eta_c$ production
for higher beam energy, suggesting a possibility
to observe $\eta_{b}$ through the diffractive productions 
by high-intensity neutrino beams available in ongoing or forthcoming
neutrino facilities.

In conclusion, we have computed the diffractive production cross sections of 
$\eta_{c,b}$ 
mesons
via the weak neutral current,
using the new results of the light-cone WFs for $Z$ and $\eta_{c,b}$.
Our results demonstrate that neutrino-induced productions will open
a new window to measure $\eta_{c,b}$.

\vspace*{0.2in}

\acknowledgements
We are pleased to acknowledge useful discussions with T. Hatsuda 
and H. Mineo in the early stage of this work.
One of the authors (A.H.) is grateful to T. Hirano for his support in numerical calculation.
A.H. is supported by JSPS Research Fellowship for Young Scientists.



\begin{references}
\bibitem{foot}
A vector dominance model \cite{BKP} gives 
$\sigma( \nu N \rightarrow \nu' N' J/\psi )
\simeq 10^{-2}$ fb for $E_{\nu}=100$ GeV with similar $E_\nu$ dependence
as our QCD result (dashed curve). At present, we do not have enough
experimental data \cite{E815} to be compared with these theoretical
predictions.
\end{references}
\end{document}